# The Paradox of Industrial Involvement in Engineering Higher Education

Srinjoy Mitra[1], Jean-Pierre Raskin[2]
[1]School of Engineering, University of Edinburgh, Edinburgh, UK
[2]School of Engineering, Université catholique de Louvain, Louvain-la-Neuve, Belgium
srinjoy.mitra@ed.ac.uk, jean-pierre.raskin@uclouvain.be

*Abstract*— This paper discusses the importance of reflective and socially conscious education in engineering schools, particularly within the EE/CS sector. While most engineering disciplines have historically aligned themselves with the demands of the technology industry, the lack of critical examination of industry practices and their impact on justice, equality, and sustainability is self-evident. Today, the for-profit engineering/technology companies, some of which are among the largest in the world, also shape the narrative of engineering education and research in universities. As engineering graduates form the largest cohorts within STEM disciplines in Western countries, they become future professionals who will work, lead, or even establish companies in this industry. Unfortunately, the curriculum within engineering education often lacks a deep understanding of social realities, an essential component of a comprehensive university education. Here we establish this unusual connection with the industry that has driven engineering higher education for several decades and its obvious negative impacts to society. We analyse this nexus and highlight the need for engineering schools to hold a more critical viewpoint. Given the wealth and power of modern technology companies, particularly in the ICT domain, questioning their techno-solutionism narrative is essential within the institutes of higher education.

## I. Introduction

Engineering as a discipline within the higher education sector enjoys an unusual status. Within the university structure, it is placed close to fundamental sciences, but more than anything else, it is a 'practice'. In that sense, Engineering is similar to Law or Medicine. While few physics graduates become physicists or geography graduates become geographers, most engineering graduates intend to remain in their discipline (e.g., >80% in Canada [1]). Like lawyers or doctors, engineering undergraduates seem to have an idea of their future profession (even if that might be changing). While actual job prospects can vary widely, depending on the country, a significant portion of engineering undergraduates do end up becoming engineers. Exact numbers are hard to determine, but this is between 40-70% in US [2] and UK [3], and certainly the highest among all STEM graduates. It is also true that engineering (including computer science and informatics) is by far the largest proportion of STEM graduates and is often higher than 60% in most Western economies [4], [5]. While we have used the term 'engineering' in a broad sense, where computer science is included, we could arguably include other applied science disciplines (e.g., applied mathematics/physics) under the same umbrella as well (i.e., all higher education that is geared to train future technologists), increasing this number even higher.

There has been a wide range of studies on why students choose an engineering education. Though these studies are mostly restricted to North-America [6], [7], we do find some key commonalities [8]. Apart from the innate interest in a field, familial pressure or job prospects (reasons that are equally present in all higher education choices), engineering students consistently mention doing social good and improving quality of life. These are also some points generally highlighted by national agencies trying to encourage more undergraduates to take up engineering as a discipline of choice. For example, the UK Royal Academy of Engineering's mission says engineering ' ….for the benefit of society', Engineers Canada mentions '…opportunity to improve lives' as the topmost reason for choosing this career [9]. Similarly, IEEE, the single largest international institution for engineers (that has links to almost all sub-branches of this discipline), also has a similar objective, '…advance technology for the benefit of humanity'. Unlike other STEM fields, engineering education is not one whose primary purpose is to know more about the universe through the lenses of that particular discipline. It's an encouragement to build something new with that knowledge, most likely for the benefit of others. However, graduate engineers will generally become part of large or small technology farms whose profit-making motives rarely match the grand ideologies of these professional bodies. This is a paradox and precisely where university-based higher education can make a difference [10].

On the other hand, engineering cohorts suffer from extreme gender inequality that becomes even wider in their professional lives as many more female graduates leave the field early [11]. Computer Science and engineering certainly have the lowest female representation among all STEM jobs (15-25% [12]). In many countries, the socio-economic diversity within an engineering cohort is also very low [12], [13]. This further enhances the nearly monolithic group of people who persist in the profession. Hence, the formative years in the university sector can be considered one of the most influential in their long professional lives. Universities education is probably one of the key opportunity to shape their world views (both existing and future) or critically confront them. Should we limit this critical period of their learning only to what the engineering industry wants? In this paper we interrogate the meaning of university-based higher education, particularly in Electrical/Electronics Engineering (EE) and Computer Science (CS) departments, when they are so entangled with the needs and ideologies of giant, for-profit technology companies.

## II. Teaching Engineering in Universities

Technology industries expect universities to create a workforce, and most of our education system is geared towards serving their vision of the future world. Some of these tech companies are not only gigantic, but they are the biggest the world has ever seen (Apple, Meta, Alphabet, Tesla, Amazon etc.), and wealthier than most governments. They are run by engineers and employ a huge number of them, but they also set the trends that other companies, industries and governments have to follow. Major ICT companies now not only decide what's good for society (e.g., as curators of public debates on digital platforms) but play a major role in upholding the power dynamics of the Global North-

South divide (e.g., by facilitating carbon emission, resource extraction and pollution) [14]. The wealth, lobbying power, and innovation capabilities of today's technology companies have pushed forward the idea that there are technological solutions to all problems humanity faces. Even if the root cause of several of them is technological 'progress' itself. This ideology is called techno-solutionism [15], and it is extremely pervasive not only within the tech industry and within engineering schools but even within liberal-minded policymakers who want to make a positive change in the most important global crisis of today [16]. Given the number of engineers the world produces (65-70% of the US STEM labour force are in engineering or computer-related occupations [17]), the role of engineers is very important concerns for the future of the planet, where technology industries and technologists play a central role. With the ever-growing complexity of engineered systems and the planetary impact of these systems '… the need for young engineers to develop nuanced understandings of the cultural, social, and political contexts…' has never been more important [18].

While questioning the techno-solutionism narrative inherent in engineering education, EE/CS disciplines probably have a unique role to play here. These are disciplines that educate students in ICET (Information Communication and Electronic Technology), on which almost all modern engineering devices and infrastructure are built. ICET advances is also one of the sources of the techno-solutionist narrative. This is primarily because of the enormous success of the tech industry in the last four-five decades that saw our lives fundamentally changed (<10% of US households had a computer in 1990, and <1% of Americans had a smartphone 20 years ago [19]). One should note that the visible ICET is just a tiny part of all the computing/communicating infrastructure that is behind almost every human-made artefact (physical or virtual) we touch today. From our global food production/supply chain to the missiles, everything is dependent on these technologies. The need to introspect the role of higher education sector in the context of creating engineers is becoming more important in EE/CS disciplines, primarily due to their omnipresence as a backbone technology in almost all industries. Yet, the negative impact (environmental and societal) of this so-called 'tech' industry is rarely taught in universities and is not easily visible to the technology researchers themselves. Nevertheless, the need for incorporating sociotechnical thinking into engineering courses has recently been voiced by several engineering education researchers [20]–[22]. The hugely important need to facilitate discussions around equality and social justice in engineering education has also been repeatedly highlighted [23], [24].

It should be noted that universities are not the only places to learn about engineering. Many European countries have 'college' systems similar to trade/vocational schools that produce a vast number of professionals. The US have also started counting such 'associates degree holders' as STEM workers who do not have a university bachelor's degree. In fact, few years of such education can be considered equally demanding and sufficient for many jobs in the technology sector. Hence, the question can even be why students choose a university-based engineering education?

III. WHAT IS ENGINEERING HIGHER EDUCATION FOR?

Engineering graduates often enter university with a grand dream of building new things that will make the world a better place. Higher education is supposed to make students more reflective and question their preconceived thoughts (better for whom? at what expense?). These are critical questions and require an understanding of justice, equality and the political/economic environment in which certain technologies are allowed to thrive. However, rather than growing more interested in these topics, engineering graduates often are found to become more disengaged from the societal implications of their profession [25]. Cech has also identified this as the 'depoliticisation' of the engineering curriculum, where there is a belief that engineering is a purely technical space, and no other dialogues should be considered seriously [18].

In the process of educating them, most educators in engineering schools (specifically in EE/CS) rely on what the tech industry needs/demands or its near-term requirements. This is not surprising since most engineering educators are also researchers in their respective domains, greatly relying on the needs of relevant industries. The reliance is not necessarily about funding (which may be the case for some individual academics), but more about the narrative of technological progress: what are the necessary and important questions worth intellectual commitment? While universities are for the public good, mostly non-profit, and should be the conscience of society, what engineering educators do is often largely governed by the ideologies of profit-making entities. In this context, engineering, as a part of university-based higher education, is not entirely unique. The competing requirements between liberal and professional education in a university have been highlighted by Newman several years ago [26]. However, professional education, as in engineering, for the service of society, was considered essential for a long period of history. Within Western nations, technological development was unquestionably assumed to associate with progress [27]. Even though this can be debated in a different sphere that looks into this success story with a decolonial lens [14], the question is whether the ideology of relentless technological development perpetuated by technology industry leaders the right thing for higher-education institutes to align with.

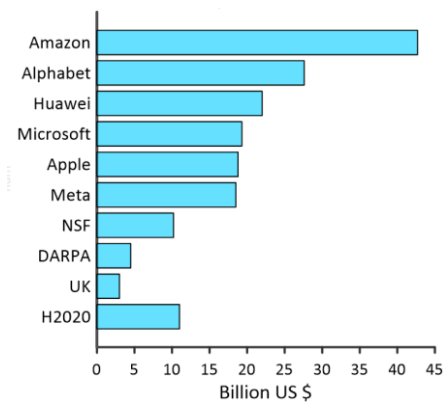

Figure 1 Approximate R&D budget of tech industry and some public funders in 2020. UK funding includes EPSRC and Innovate UK, H2020 shows the average budget over 7 year period. (Source [30], [35], [36]).

As pointed out before, ICET innovation has shown unprecedented 'success' and continues to be the largest sector in terms of R&D expense. But today, it is not the governments who are funding the bulk of the research anymore, at least in the US and EU. The wealthiest companies surpassed that threshold several years ago [28]. Even the research powerhouse of the US government funding looks hugely unimpressive compared to the tech behemoths of today (see Fig.1). Though government funding still fuels fundamental discoveries in (applied) sciences [29], the unprecedented cycle of innovation and funding in the ICET domain lets the technology companies project their importance within the public sphere including the academic circles. These figures don't show the entire reality of funding in engineering research. While the government funding in many cases includes fundamental sciences (and is difficult to separate from engineering/informatics) [30], it can be safely assumed the R&D expenses of these tech firms are mostly on ICET itself. Several other tech companies, with huge R&D budget often spend much more on a single project than several Global-North governments can do in decades (e.g., nearly $100bn on Metaverse [31] or billions invested in the Blue Origin space project [32]).

A large chunk of the industrial research budget goes to universities as well, particularly the ones with strong CS/EE discipline (close to 15% of the entire research budget in some top US universities [33]). It is not surprising to understand why the engineering schools are so indebted to these companies, and eager to hear from their university advisory officers [18]. It would be quite unusual to see this in other parts of a standard university (except maybe in the business school). For example, if a School of Geoscience has close contact with the fossil fuel industry or a School of Neuroscience depends hugely on the gaming industry's funding/input, there will be a significant pushback from within. Even within business schools there is an ongoing debate on what the overall alignment of a higher education institute to profit-making entities should be [34]. Unlike other faculties, unfortunately, there is a substantial lack of such criticism within engineering/computing schools or even within the community focused on engineering education.

It should be noted that this is not a question of conflict of interest around an individual academic's own research funding. There has been long debates about academic research (e.g., in pharmaceutical, nutrition) where various companies directly benefit from specific results that has also been funded by them [37]. The issue we highlight here goes much beyond that. In fact, only a very few high-profile academics get large amounts of the gigantic tech industry research budget. The disciplinary allegiance to the narrative of techno-solutionism is not a matter of individual academic's (mis)judgement. It has gone far beyond this and become an ingrained ideology within entire engineering schools. Given engineering educators largely rely on their prior experience, instincts (and peer pressure) to judge what is important in the curriculum, it's important to call this into question. The standardisation organisations in different countries do try to feed in the need for some social factors (e.g., engineering ethics), but they mostly stick to codes of practice and distinctively lack in the questions of social justice [38]. Unfortunately, there is rarely any discussion at all relating to global justice in engineering education. In the last ten years, the European Journal of Engineering Education has 'industry' in the title of more than 70 publications, but just 3 with 'justice' in it. Though a detailed analysis is outside the scope of this paper, but we can assume that very few of these publications will associate negative characterisation of the engineering industry. On the other hand, the lack of mention of 'justice' in the title, keywords or even in the text is a significant omission.

Hence in the four to six years of university education in engineering, students will rarely encounter such concepts of social and global justice. As Riley and others have demonstrated, 'critical thinking' (CT) within engineering curriculum is more about problem solving, and applied within focused elements of the topic 'but not *about* engineering itself' [39]. The conflict between the 'instrumental' and 'emancipatory' demands of higher education is probably most visible in the engineering departments in a university [40]. With the omnipresence of ICET in almost all aspects of engineering, we consider this EE/CS departments should be more acutely aware of their responsibilities to the society.

ICET products and services depend on critical resources and cheap labour, often sourced from exploiting the Global South. This colonial legacy of the discipline and a debate on the never-ending resource extraction (even for the 'greenest' of technologies) is generally not a part of the curriculum [41]. CS/EE graduates would most likely come across futuristic technological solution for the climate crisis (e.g., using AI or blockchain), and asked to develop similar solutions. However, the incredible environmental impact of our never-ending demand for faster computing, higher data-bandwidth and more memory will rarely be part of the education. As the industry demands, engineering educators need to cram constantly increasing technological material to keep their courses relevant. There is hardly any time to reflect why something is being taught at all. As Cech and Sherick points out, a typical course of undergrad engineering '…not only reflects the ideology of depoliticisation, it also reinforces it.'[18]. Hence, they suggest to reduce technical content from engineering curriculum!

Even if the funding mechanism of universities varies widely (around the world and even within a country), it can be agreed that universities gather prestige from their public face. By educating responsible citizens of the coming generation, doing research for public good, publishing in academic journals (largely supported by free labor of publicly funded researchers), and generally being a non-profit entity with philanthropic roots. We should note that most large universities include non-STEM departments as well, and their understanding of a university's reputation mostly relies on the overall intellectual environment rather than access to industrial funding. So, why would the (mostly) public-funded higher education institutes serve the need (in terms of producing work force and knowledge) for a technocratic elite whose intentions are not aligned with the benefit of humanity? If industries need specific skills (which might anyway get outdated in a few years) in their employees, they should provide it themselves. In contrast, universities should produce graduates who can critically question the job they are doing, beyond the requirement that is imposed by their job profile. Hence, replacing technological content with more techno-social understanding should probably be a key goal of engineering higher education.

IV. GRADUATES AND ESTEEMED CONFERENCES

Universities generate not only professional graduate engineers but also future educators. In fact, the number of Masters students in STEM (in the US) is again skewed towards EE/CS (nearly double compared to rest of STEM) and similarly high for doctoral degrees awarded [3], [5]. Though these numbers are from US and UK, they provide a representative sample and demonstrate the dominance of ICET researchers in STEM higher education. A significant section of these people will not only take faculty positions but will also have more decision-making positions in engineering research. These 'early career researchers' (masters and PhDs) regularly write conference (and journal) papers, and visit these scholarly gatherings. Being accepted to some of the most well-known technology conferences is not just a learning opportunity but can also make or break one's career, particularly if they want to get into academia. Within the EE/CS domain these conferences are mainly dominated by scholarly communities like IEEE and ACM (Association for Computing Machinery). Students (and their supervisors) spend months gathering data and writing papers for some of the most prestigious conferences (ISSCC: International Solid State Circuits Conference, IEDM: International Electron Devices Meeting, DATE: Design, Automation and Test in Europe, CCS: Computer and Communications Security etc.). The importance of these and many other smaller conferences for academic progress of students and for creating collaboration opportunities cannot be overstated.

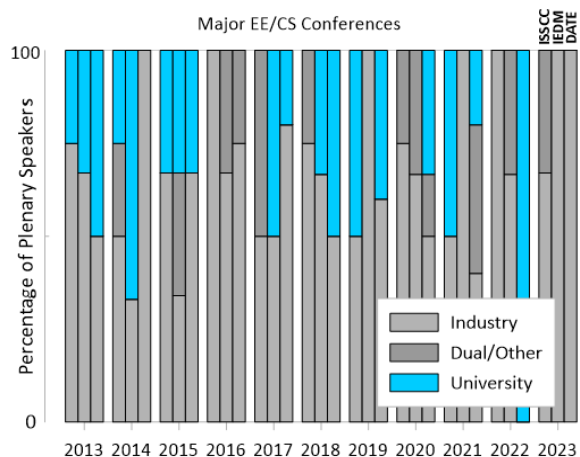

Figure 2 Plenary/keynote speaker affiliation in three high-profile EE/CS conferences. Speakers with dual affiliation (i.e., university and industry) or from non-university research are marked as 'Dual/Other'. Not all types of speakers were present in all years. (Source: Individual conference websites)

However, while academics and students will primarily dominate most similar scholarly conferences in many other fields, one can see a huge presence of industry in high-profile engineering conferences. Some conference papers will be jointly written by researchers in industry and universities or sometimes entirely by industry-based researchers. This in itself is not a problem other than demonstrating the same point that industrial funding is hugely important for these higher education scholars. However, we can also notice the unique importance given to leadership from tech industries in the increasing number of plenary/keynote speeches they are invited to. To be clear, these conferences are not the ones that are primarily focused on the tech industry to showcase their products (e.g., Consumer Electronics Show, Mobile World Congress), and attended mostly by the industry itself. A typical IEEE/ACM conference will be attended mostly by graduate students and university-based researchers. However, close to 75% of plenary/keynote speeches on average (see Fig. 2) are given by 'visionaries' who hold key positions in the tech industry. Hence, it is rather obvious that the dreams and achievements of the industry sets the tone of what's important in this domain. This gives graduates (and future educators) a very specific sense of what is considered serious research in their discipline and worth spending their intellectual time on. Not surprisingly, creating more technology is most often this industry's prescribed solution for all illnesses. Even when the speakers are all from academia, industrial sponsorship often plays a huge part in hosting many of these high-profile conferences. This trend of providing a platform to the industry and its leaders within an otherwise academic conference is not new and can be traced back several years. What do industry leaders get out of this is rarely questioned. It is possible that they need to validate their visionary ideas of the future world with people who supposedly hold the authority on what's truly worthy, i.e., the academic community.

## V. How does it matter?

Why would such a deep nexus between engineering education and industry be a bad idea? The phenomena described here probably started from Stanford's School of Engineering dean Frederick Turman who aligned the school's research policy with military priorities and the budding tech industry [42]. This created a constant knowledge flow between university and industry, and also earned him the nick name 'Father of Silicon Valley'. The model was soon copied by other US universities and got exported to Europe. Though this might have resulted in significant achievements and human ingenuity in the ICET sector, it also placed such companies as the flagbearer of technological progress.

The reach of modern technology firms is much more than we often appreciate. The global chip shortage of 2020/21 has shown how much the world depends on just one facet of the ICET sector (the silicon chip) [43] and how within a few months that eventually led to serious geopolitical interventions [44]. Hundreds of billions of dollars have suddenly been invested in US and EU to achieve global dominance in silicon chips. Similar companies and associated high-tech startups are also the key drivers of the green-tech culture. Ideas like eVTOL (Vertical Take-off and Landing aircraft), CCS (carbon capture and storage), geoengineering etc. are all technological solutions to climate crisis that receive enormous funding from private and public sectors. However, who makes sure that these technologies have any realistic possibility for a truly sustainable world? Ecological economists have consistently shown that none of these solutions can be built to scale for a meaningful change without creating more ecological damage elsewhere in the world [45]. Afterall, it is now widely known that the carbon emission from ICT sector alone is more than that of the aviation sector, and is growing every year with no sign of limits [46]. On top of that, most of these technologies require enormous resource extraction from (e.g., rare earth minerals) from the most vulnerable regions of the word [47]. Even the med-tech industry, with their benevolent goals of improving healthcare, do not need to confront the possibility that more healthcare technology rarely equates to more healthcare access, and often diverts resources from the poor [48]. It is not enough that the criticism to such unquestioned technological excellence come from other disciplines within academia, the engineering schools should play a front and center role in this reflection.

Today, with the climate crisis at the doorstep of the Global North, the need for 'sustainability' is everywhere. Sustainable Development Goals (SDG) is a target for almost all governments and institutions. Universities are increasingly addressing sustainability by introducing various travel policies, printing regulations or encouraging recycling within campus [49]. Though it is unclear how effective these measures are being, they are all commendable. However, one could argue that the curriculum of their EE/CS schools can play one of the biggest impacts in any sustainability activity. If students are exposed to a world view primarily driven by technology leaders and their understanding of global order and sustainability, there is hardly any chance of debate. This ideology of techno-solutionism is primarily to preserve the technocratic dominance of the public sphere that we accepted for a long time. In their book *Power and Progress* [50] Jonson and Acemolglu explains how this dominant narrative of equating technological development with civilisational progress is a convenient myth propagated by industry and often with the help from academia [51].

## VI. ICET Education for the Future

Leaving aside such arguments, even if the success of the tech industry in shaping a world view is purely an emergent property; there is considerable danger in considering net-zero, and SDG goals can be achieved largely by technological means, without the need to give up anything substantial [52]. The utopic techno-solutionist idea suggests that this will happen sometime in the future without considering the ongoing climate injustice that the Global South faces right today. Given that the climate-crisis and planetary boundaries are some of the most important aspects we should bw all aware of, ICET educators need to confront them right now. Given the size of the industry and their R&D budgets, the impact of EE/CS graduates in the future world is obvious. The ideology and fetishisation of technological innovation has already seen its limit in climate crisis and neocolonial power structure. Since tech industries are by far the largest commercial ventures in the entire world, they play a prominent part in not only exacerbating the climate crisis but perpetuating the techno-solutionist narrative [53]. This needs to be questioned from within the university educational system. At least SDG should be at the forefront of engineering education, then anything else. Sustainability and climate mitigation are probably the most important things all graduates should debate now and more importantly in engineering. Similar suggestions has been made before [54], [55], but without the emphasis on EE/CS streams.

We call for less intervention from industry and more from the disciplinary breadth of the university itself: reflective, intellectually broadening, higher-education should be a required aspect of the entire techno-scientific curriculum including engineering. This could mean some undergraduates would decide not to do engineering in the end. Maybe they have understood how to better use their intellectual effort for a more equitable and sustainable planet for everyone.

## VII. Conclusion

Here we interrogate the problem regarding unquestioned influence of the technology industry in university-based engineering higher education. We demonstrate this rather obvious connection using indirect data on research funding dependency and the platform given to industrial leadership. One important question probably is why should we care? It is not because of the (supposed) dilution of the intellectual sphere within the university from self-interested, profit-making entities. But it is more because of the dominance of the technology sector in our everyday lives and the proliferation of ICET professionals every year. How can we make sure that these students are educated to confront the most important challenges of tomorrow in a just and equitable way. Accreditation bodies can play a part, but accreditation evaluators are usually practicing engineers from industry and academia; few are formally trained in the socio-cultural contexts of technology. The change should come from the within and by interrogating the academic-industry nexus. Sustainability and climate crisis are not something we can wait for years to be solved (by the tech industry), while the relentless extraction and exploitation of the Global South continues.

Finally, we come back to the question of the unique position of engineering in a university system. Other practice-driven education such as Law and Medicine will produce graduates who mostly continue doing what they learned in the university without a huge demand for innovation. Most doctors will practice medicine by the book and lawyers follow existing legal codes. However, a huge part of technology industry's demand is to make something new, to make profit out of tools and services that are 'better' than what was there before. The need for innovation is fundamental to the capitalist economy they are part of. Hence their judgement for what to innovate for whose good and who gets left behind (or negatively impacted) is hugely important.

ACKNOWLEDGMENT

The authors acknowledge fruitful discussions with colleagues, particularly Prof Adrian Friday, Lancaster University.